\documentclass[copyright,creativecommons]{eptcs}

\usepackage{amsmath,amsfonts,amssymb,amsthm,stmaryrd}
\usepackage{branesymb,proof,enumitem}

\DeclareMathAlphabet{\mathpzc}{OT1}{pzc}{m}{it}

\newcommand{\N}{\mathbb{N}}
\newcommand{\Q}{\mathbb{Q}}
\newcommand{\R}{\mathbb{R}}

\newcommand{\A}{\mathbb{A}}
\newcommand{\T}{\mathbb{T}}
\renewcommand{\P}{\mathbb{P}}
\newcommand{\M}{\mathbb{M}}

\newcommand{\set}[1]{\{ #1 \}}
\newcommand{\trs}{\xrightarrow}
\newcommand{\rlabel}[1]{\; \text{\scriptsize{(#1)}}}
\newcommand{\infr}[3]{\infer[\text{\scriptsize{(#1)}}]{ #2 }{ #3 }}
\newcommand{\func}[1]{\llbracket #1 \rrbracket}
\newcommand{\Mid}{\; \mid \;}

\newcommand{\mpar}[2]{{}_{#1}\kern-2pt\varobar_{#2}}
\newcommand{\mtimes}[2]{{}_{#1}\kern-1pt\otimes_{#2}}
\newcommand{\mAt}[2]{@^{#1}_{#2}}

\newcommand{\sub}[1]{\begin{subarray}{c} #1 \end{subarray}}

\newcommand{\sys}{\mathsf{sys}}
\newcommand{\mem}{\mathsf{mem}}
\newcommand{\act}{\mathsf{act}}

\newcommand{\mset}{\mathpzc}

{\medskip\par\noindent%
\addtolength{\abovedisplayskip}{-0.3\abovedisplayskip}%
\addtolength{\belowdisplayskip}{-0.6\belowdisplayskip}%
\rule{1pt}{4pt}\leaders\hrule depth-3pt height4pt\hfill%
~\textsf{\textbf{#1}}~%
\leaders\hrule depth-3pt height4pt\hfill\rule{1pt}{4pt}%
\nopagebreak}%
{\nopagebreak\rule{1pt}{4pt}\leaders\hrule height1pt\hfill\rule{1pt}{4pt}
\newline}

\newtheoremstyle{theorem}
{\topsep}
{\topsep}
{\it}
{}
{\bfseries}
{.}
{0.5em}
{\thmname{#1}\thmnumber{ #2}\thmnote{ (#3)}}%

\theoremstyle{theorem} 
\newtheorem{theorem}{Theorem}[section]
\newtheorem{proposition}[theorem]{Proposition}
\newtheorem{lemma}[theorem]{Lemma}
\newtheorem{corollary}[theorem]{Corollary}
\newtheorem{definition}[theorem]{Definition}
\newtheorem{example}[theorem]{Example}

\title{Measurable Stochastics for Brane Calculus%
\footnote{Work funded by MIUR PRIN project ``SisteR'',
   prot.~20088HXMYN.}}
\author{Giorgio Bacci \qquad\qquad\qquad Marino Miculan
\institute{Department of Mathematics and Computer Science, University of Udine, Italy}
\email{giorgio.bacci@uniud.it \qquad\qquad marino.miculan@uniud.it}
}
\def\titlerunning{Measurable Stochastics for Brane Calculus}
\def\authorrunning{G. Bacci \& M. Miculan}

\begin{document}
\maketitle

\begin{abstract}
  We give a stochastic extension of the Brane Calculus, along the
  lines of recent work by Cardelli and Mardare \cite{cm:quest10}.  In
  this presentation, the semantics of a Brane process is a measure of
  the stochastic distribution of possible derivations.  To this end,
  we first introduce a labelled transition system for Brane Calculus,
  proving its adequacy w.r.t.~the usual reduction semantics.  Then,
  brane systems are presented as Markov processes over the measurable
  space generated by terms up-to syntactic congruence, and where the
  measures are indexed by the actions of this new LTS.  Finally, we
  provide a SOS presentation of this stochastic semantics, which is
  compositional and syntax-driven.
\end{abstract}

\section{Introduction}

The Brane Calculus (BC) \cite{cardelli04:bc} is a calculus of mobile
processes designed for modeling membrane interactions within a cell.
A process of this calculus represents a system of nested membranes,
carrying their active components \emph{on} membranes, not inside them.
Membranes interact according to three reaction rules, corresponding to
phagocytosis, endo/exocytosis, and pinocytosis.

In the original definition, reaction rules do not consider
quantitative aspects like rates, volumes, etc.  However, it is
important to address these aspects, e.g. for implementing stochastic
simulations, or for connecting Brane Calculus with quantitative models
at lower abstraction levels (such as stochastic $\pi$-calculus and
$\kappa$-calculus for protein interactions).

In this paper, we introduce a stochastic semantics for the Brane
Calculus.  Clearly, a stochastic calculus could be obtained just by
adding rates to reaction rules; however, the resulting ``pointwise''
rated reduction semantics is not fully satisfactory for several
reasons.  First, it is not compositional, i.e., reaction rates of a
process are not given in terms of the rates of its components.
Secondly, stochastic reaction rules are not easy to deal with in
presence of large populations of agents (as it is often the case in
biological systems), because we have to count large number of
occurrences for calculating the effective reaction rates. Third, it
does not generalize easily to other quantitative (e.g. geometric)
aspects.

To overcome these issues, we adopt a novel approach recently
introduced by Cardelli and Mardare \cite{cm:quest10}, which is
particularly suited when a measure of similarity of behaviours is
important (similar ideas have been proposed for probabilistic automata
\cite{kwiatkowska:arts99, sl:njc95}, and Markov processes
\cite{hermanns02, bhk:YMCA06, prakashbook09}).  The main point of this
approach is that the semantics of a process is a \emph{measure} of the
stochastic distribution of the possible outcomes.  Thus, processes
form a \emph{measurable space}, and each process is given an
action-indexed family of measures on this space.  For an action $a$,
the measure $\mu_a$ associated to a process $P$ specifies for each
measurable set $S$ of processes, the rate $\mu_a(S)\in \R^+$ of
$a$-transitions from $P$ to (elements of) $S$.  The resulting
structures, called \emph{Markov processes} (MPs),
are not continuous-time Markov chains because each transition is from
a state to a possibly infinite class of states (closed to the
congruence relation over processes) and consequently cannot be
described in a pointwise style.  An advantage of this approach is that
we can apply results from measure theory for solving otherwise
difficult issues, like instance-counting problems; moreover, process
measures are defined compositionally, and can be characterized also by
means of operational semantics in GSOS form. Finally, other measurable
aspects of processes (e.g., volumes) can be dealt with along the same
lines.

However, the approach of \cite{cm:quest10} has been applied to CCS
only, and in order to adapt it to BC, we have to solve some problems.
First, in order to define systems of BC as Markov processes, we need
to define \emph{actions} for brane systems, which corresponds to
define a \emph{labelled transition system} (LTS) for BC.  Defining
correctly a labelled transition system for a complex calculus like BC
is notoriously a difficult task, because labels should describe
precisely how a system can interact with the surrounding environment.
In our case, we have been inspired by the so-called \emph{IPO
  construction} \cite{lm:concur00}, following the approach of
\cite{rs:concur08,bgm:fossacs09} for the case of Mobile Ambients.
As a result, in Section~\ref{sec:lts} we introduce the first labelled
transition system for BC, and prove its adequacy with respect to the
original reduction semantics.

A peculiar feature of this LTS is that after a transition, a system
can yield a \emph{higher-order} term, i.e., a system with ``holes''
(like $\pi$-calculus' ``abstractions'').  This has several
consequences on our work.  First, we have to define a suitable syntax
for these higher-order terms; for this reason, in Section~\ref{sec:bc}
we introduce a \emph{simply typed} version of brane calculus from the
beginning, extended with metavariables, $\lambda$-abstractions and
applications.  Well-formedness of terms is guaranteed by a suitable
typing system; e.g., membranes and systems are represented by ground
terms of type $\mem$ and $\sys$, respectively.  Secondly, the
bisimulation definition has to be accommodated in order to deal with
transitions yielding higher-order terms (Section~\ref{sec:lts}).
Finally, also the definition of Markov kernel is affected, as we will
see in Section~\ref{sec:sbc}: the kernel cannot be defined simply on
the space of brane systems, but must consider also these higher-order
terms---in fact, the measure will be defined over all well-typed
terms, also those higher-order.

After that these issues have been addressed and the Markov kernel has
been defined, we can look for a simpler presentation of the semantics
of Markov processes. In Section~\ref{sec:bisim} we present a SOS
system for processes, capturing the Markov kernel over processes: the
stochastic bisimilarity induced by this SOS semantics corresponds to
the Markov bisimilarity defined in Section~\ref{sec:sbc}.  Therefore,
this semantics can be fruitfully used for simulations, or for
verifying system equivalences.

Some concluding remarks and directions for further work are in Section~\ref{sec:concl}.

\section{Brane Calculus}
\label{sec:bc}

In this section we recall Cardelli's Brane Calculus
\cite{cardelli04:bc} focusing on its basic and finite version (without
communication primitives, molecular complexes and replication).  Here
we adopt an alternative presentation of the calculus: instead of the
traditional term grammar of \cite{cardelli04:bc}, we introduce an
unstructured version but equipped with a type system.  Although typed
terms may seem unnecessary now, they will be useful in
Section~\ref{sec:lts}, where a labelled transition semantics will be
introduced.

\paragraph{Syntax}
\label{sec:syntax}

The grammars for terms and types are specified below.
\begin{align*}
(Terms) && M &\mathrel{::=} \zero \Mid \void \Mid X \Mid \alpha.M \Mid 
	M | M \Mid M \comp M \Mid \cell{M}{M} \\
 && \alpha &\mathrel{::=} \phago_{n} \Mid \cophago_{n}(M) \Mid
	\exo_{n} \Mid \coexo_{n} \mid \pino_{n}(M)
\\[0.5ex]
(Types) && t &\mathrel{::=} \sys \Mid \mem \Mid \act
\end{align*}
The subscripted names $n$ are taken from a countable set 
$\Lambda$, while term variables $X$ are taken from 
a countable set $\mathcal{X}$, assumed to be disjoint from each other.

\looseness=-1
Types are assigned to terms as usual. 
A type \emph{environment} $\Gamma$ is a finite map from 
term variables to types. If $\Gamma$ is an environment and $X$ a 
variable not in the domain of $\Gamma$, we denote by 
$\Gamma, X : t$ the environment which assigns type $\Gamma(Y)$ to 
each variable $Y \in \mathit{dom}(\Gamma)$ and type $t$ to $X$;
if $\Gamma_{1}$ and $\Gamma_{2}$ have disjoint domains, 
$\Gamma_{1}, \Gamma_{2}$ denotes the environment which assigns 
type $\Gamma_{1}(X)$ to variables $X \in \mathit{dom}(\Gamma_{1})$
and $\Gamma_{2}(Y)$ to variables $Y \in \mathit{dom}(\Gamma_{2})$.
The type inference rules are given in Table~\ref{tbl:typesystem}.
\begin{table}[t]
\hrule
\begin{gather*}
\rl{\Gamma \vdash X : t}{\Gamma(X) = t}
\rlabel{var}
\qquad
\rl{\Gamma \vdash a : \act}{
  a \in \set{\phago_{n}, \exo_{n}, \coexo_{n}}}
\rlabel{act}
\qquad
\rl{\Gamma \vdash a(M) : \act}{
  a \in \set{\cophago_{n}, \pino_{n}}
  \quad
  \Gamma \vdash M : \mem}
\rlabel{act-arg}
\\[1ex]
\rl{\Gamma \vdash \zero : \mem}{}
\rlabel{zero}
\qquad
\rl{\Gamma_{1}, \Gamma_{2} \vdash \alpha.M : \mem}{
  \Gamma_{1} \vdash \alpha : \act
  \quad
  \Gamma_{2} \vdash M : \mem}
\rlabel{$\alpha$-pref}
\qquad
\rl{\Gamma_{1}, \Gamma_{2} \vdash M | N : \mem}{
  \Gamma_{1} \vdash M : \mem
  \quad
  \Gamma_{2} \vdash N : \mem}
\rlabel{par}
\\[1ex]
\rl{\Gamma \vdash \void : \sys}{}
\rlabel{void}
\qquad
\rl{\Gamma_{1}, \Gamma_{2} \vdash \cell{M}{N}: \sys}{
  \Gamma_{1} \vdash M : \mem
  \quad
  \Gamma_{2} \vdash N : \sys}
\rlabel{loc}
\qquad
\rl{\Gamma_{1}, \Gamma_{2} \vdash M \comp N : \sys}{
  \Gamma_{1} \vdash M : \sys
  \quad
  \Gamma_{2} \vdash N : \sys}
\rlabel{comp}
\end{gather*}
\hrule
  \caption{Typing system for Brane Calculus}
  \label{tbl:typesystem}
\end{table}
Notice that this type system admits only \emph{linear} terms, that is, each
variable can occur at most once. Indeed, in rules ($\alpha$-pref),
(par), (loc), and (comp), the environment extension $\Gamma_{1},
\Gamma_{2}$ is defined only when $\Gamma_{1}$ and $\Gamma_{2}$ have
disjoint domains.

In the rest of the paper we assume to work only with well-typed terms.
The set of well-typed terms will be denoted as $\T$, while $\P$ and
$\M$ denote the set of terms of type $\sys$ and $\mem$, respectively.
By convention we shall use $x$, $y$ \dots\ for variables of type
$\mem$, and $X$, $Y$ \dots\ for variables of type $\sys$.  A similar
convention is used for base type terms: $\sigma$, $\tau$, \dots\ are
terms in $\M$, while $P$, $Q$, \dots\ are terms in $\P$.

Systems can be rearranged according to a structural congruence
relation; the intended meaning is that two congruent
terms actually denote the same system.
Structural congruence, $\equiv$, is the smallest equivalence relation 
on (possibly open) terms of the language which contains the axioms 
and rules listed below. 
We write $\Gamma \vdash M \equiv N$ as shorthand for 
$\Gamma \vdash M : t$ and $\Gamma \vdash N : t$ and $M \equiv N$.
Notice that this notation implicitly assumes that structural equivalent
terms must be of the same type.
\begin{gather*}
\Gamma \vdash P\comp Q \equiv Q \comp P \qquad
\Gamma \vdash P\comp(Q\comp R) \equiv (P\comp Q)\comp R \qquad
\Gamma \vdash P\comp \void \equiv P   \qquad
\Gamma \vdash \cell{\zero}{\void} \equiv \void \qquad
\\[1ex]
\Gamma \vdash \sigma|\tau \equiv \tau | \sigma \qquad
\Gamma \vdash \sigma |(\tau | \rho) \equiv (\sigma | \tau) | \rho \qquad
\Gamma \vdash \sigma | \zero \equiv \sigma 
\\[1ex]
\rl{\Gamma \vdash P \comp R \equiv Q \comp R}{\Gamma \vdash P \equiv Q} \qquad
\rl{\Gamma \vdash \sigma | \rho \equiv \tau | \rho}{\Gamma \vdash \sigma \equiv \tau} \qquad
\rl{\Gamma \vdash \cell{\sigma}{P} \equiv \cell{\tau}{Q}}{\Gamma \vdash P \equiv Q  \quad \Gamma \vdash \sigma \equiv \tau} \qquad
\\[1ex]
\rl{\Gamma \vdash \alpha.\sigma \equiv \beta.\tau}{\Gamma \vdash \alpha \equiv \beta \quad \Gamma \vdash \sigma \equiv \tau} \qquad
\rl{\Gamma \vdash \pino(\sigma) \equiv \pino(\tau)}{\Gamma \vdash \sigma \equiv \tau} \qquad
\rl{\Gamma \vdash \cophago_n(\sigma) \equiv \cophago_n(\tau)}{\Gamma \vdash \sigma \equiv \tau} \qquad
\end{gather*}
With respect to the structural congruence of \cite{cardelli04:bc}, we
have added the possibility of rearranging the sub-membranes contained
in co-phago and pino actions (the last three rules of the table above).

\paragraph{Reduction Semantics}
\label{sec:redsemantics}
The dynamic behaviour of Brane Calculus is specified by means of a reduction semantics, defined by means of a reduction relation (``reaction'') 
$\react \subseteq \P \times \P$, whose rules are listed in 
Table~\ref{tbl:redrules}.
\begin{table}[t]
\hrule
\begin{gather*}
\rl{\cell{\cophago_n(\rho).\tau | \tau_0}{Q} \comp \cell{\phago_n.\sigma | \sigma_0}{P} \react \cell{\tau | \tau_0}{\cell{\rho}{\cell{\sigma | \sigma_0}{P}} \comp Q}}{}
\rlabel{red-phago}
\\[1ex]
\rl{\cell{\coexo_n.\tau | \tau_0}{\cell{\exo_n.\sigma | \sigma_0}{P} \comp Q} \react \cell{\sigma | \sigma_0 | \tau | \tau_0}{Q} \comp P}{}
\rlabel{red-exo}
\\[1ex]
\rl{\cell{\pino(\rho).\sigma | \sigma_0}{P} \react \cell{\sigma | \sigma_0}{\cell{\rho}{\void} \comp P}}{}
\rlabel{red-pino}
\\[1ex]
\rl{\cell{\sigma}{P} \react \cell{\sigma}{Q}}{P \react Q}
\rlabel{red-loc}
\qquad
\rl{P \comp R \react Q \comp R}{P \react Q}
\rlabel{red-comp}
\qquad
\rl{P \react Q}{P \equiv P' \quad P' \react Q' \quad Q'\equiv Q}
\rlabel{red-equiv}
\end{gather*}
\hrule
\label{tbl:redrules}
\caption{Reduction semantics for the Brane Calculus.}
\end{table}
It is easy to see that subject reduction holds. Note that the presence of
(red-equiv) makes this not a structural presentation, since term structure
can change according with $\equiv$.

\section{A Labelled Transition Semantics for Brane Calculus}
\label{sec:lts}

In this section we introduce a labeled transition system (LTS) for the
Brane calculus, along the lines of \cite{rs:concur08} where a LTS is
outlined in the case of Mobile Ambients. A SOS style presentation of the semantics will help in the
definition of the stochastic semantics of Section~\ref{sec:sbc}.

We shall use a meta-syntax for simple syntactic manipulation of terms.
The meta-syntax is a simply typed $\lambda$-calculus and can be
thought as a primitive system of higher order abstract syntax.  We
extend the base types with function types, in order to type terms in
the meta-syntax.  We also add to the term signature the
$\lambda$-abstraction and application constructs, which should not be
considered as a language extension: their function is to allow for a
structural definition of a labelled transition system, and as such
they have no computational meaning.
\begin{align*}
M \mathrel{::=} \dots \Mid \lambda X : t.\; M \Mid M(M)
& & 
t \mathrel{::=} \dots \Mid t \to t
\end{align*}
The $\lambda$-abstraction binds variables. Terms are taken up-to
$\alpha$-equivalence on bound variables.

Terms in the meta-language are typed using the standard typing rules for
typed $\lambda$-terms, and the two rules below are added to the set of
of type rules of Table~\ref{tbl:typesystem}.
\begin{equation*}
\rl{\Gamma \vdash \lambda X : t. \; M : t \to t'}{
  \Gamma, X:t \vdash M : t'}
\rlabel{lambda}
\qquad
\qquad
\rl{\Gamma \vdash M(N) : t'}{
  \Gamma \vdash M : t \to t'
  \qquad
  \Gamma \vdash N : t}
\rlabel{app}
\end{equation*}

We consider terms to be syntactically equal up-to
$\beta\eta$-equivalence, i.e., the smallest congruence that contains
the two axioms $(\lambda X : t.\; M)(N) = M\set{N/X}$ and $\lambda X :
t.\; M(X) = M$.  Thus, for instance, $(\lambda X:\sys.\;
\cell{\sigma}{X})(P)$ and $\cell{\sigma}{P}$ are the same term.  In
the remainder of the paper, when we write a meta-syntax term of base
type ($\sys$, $\mem$ or $\act$), we mean the term obtained by complete
$\beta\eta$-reduction (which exists because our metalanguage is a
simply typed-$\lambda$ calculus \cite{baren:lcalc}).
This assumption will ease the description avoiding many technicalities
which are out of the scope of the paper.

We extend structural congruence to meta-syntactic terms by adding 
the following rule.
\begin{gather*}
\rl{\Gamma \vdash \lambda X : t.\; P \equiv \lambda X : t.\; Q}{
  \Gamma, X:t \vdash P \equiv Q}
\end{gather*}
Structural congruence is compatible with $\lambda$-terms evaluation 
in the following sense:
\begin{lemma}
If $M \equiv N : t \to t'$ and $T \equiv S : t$
then it follows that $M(T) \equiv N(S) : t'$.
\end{lemma}

\looseness=-1
We can now define a labelled transition system for the Brane Calculus,
following \cite{rs:concur08}.  This system will be structural (i.e.,
in SOS format) and finitary branching; these features will allow to
define smoothly the stochastic semantics in Section~\ref{sec:sbc}.  To
our knowledge this is the first finitary structural operational
semantics for the Brane Calculus. (An LTS for Brane Calculus has been
given in \cite{bmmt:08}, but it was not structural nor finitary
branching.)  The rules of our LTS are given in Table~\ref{tbl:lts},
and are organized into two parts: rules for membrane terms and for
system terms.  In the definition we have omitted types since the
upper/lower case notation for $\sys$/$\mem$ typed variables makes the
presentation clearer. The only exception to this notation is in
(phago) and (L/R$\comp$-phago) rules, where the variable $Z$ has type
$(\sys \to \sys)$.

\begin{table}[t]
\hrule
\begin{gather*}
\rl{ \phago_{n}.\sigma \trs{\phago_{n}} \sigma}{}
\rlabel{$\phago$-pref}
\qquad
\rl{ \cophago_{n}(\rho).\sigma \trs{\cophago_{n}(\rho)} \sigma}{}
\rlabel{$\cophago$-pref}
\qquad
\rl{ \pino_{n}(\rho).\sigma \trs{\pino_{n}(\rho)} \sigma}{}
\rlabel{$\pino$-pref}
\\[0.5ex]
\rl{ \exo_{n}.\sigma \trs{\exo_{n}} \sigma}{}
\rlabel{$\exo$-pref}
\qquad
\rl{ \coexo_{n}.\sigma \trs{\coexo_{n}} \sigma}{}
\rlabel{$\coexo$-pref}
\qquad
\rl{ \sigma | \tau \trs{\alpha} \sigma' | \tau}{ \sigma \trs{\alpha} \sigma' }
\rlabel{L-par}
\qquad
\rl{ \tau | \sigma \trs{\alpha} \tau | \sigma' }{ \sigma \trs{\alpha} \sigma' }
\rlabel{R-par}
\\[0.5ex]
\rl{ \cell{\sigma}{P} \trs{\text{phago}_{n}} \lambda Z.\; Z(\cell{\sigma'}{P}) }{ \sigma \trs{\phago_{n}} \sigma'}
\rlabel{phago}
\qquad
\qquad
\rl{ P \comp Q \trs{\text{phago}_{n}} \lambda Z.\; (F(Z) \comp Q) }{ P \trs{\text{phago}_{n}} F }
\rlabel{L$\comp$-phago}
\\[0.5ex]
\rl{ \cell{\sigma}{P} \trs{\overline{\text{phago}}_{n}} \lambda X.\; \cell{\sigma'}{ \cell{\rho}{X} \comp P} }{ \sigma \trs{\cophago_{n}(\rho)} \sigma'}
\rlabel{$\overline{\text{phago}}$}
\qquad
\qquad
\rl{ P \comp Q \trs{\overline{\text{phago}}_{n}} \lambda X.\; (A(X) \comp Q) }{ P \trs{\overline{\text{phago}}_{n}} A }
\rlabel{L$\comp$-$\overline{\text{phago}}$}
\\[0.5ex]
\rl{ \cell{\sigma}{P} \trs{\text{exo}_{n}} \lambda Xy.\; \cell{\sigma' | y}{X} \comp P }{ \sigma \trs{\exo_{n}} \sigma'}
\rlabel{exo}
\qquad
\qquad
\rl{ P \comp Q \trs{\text{exo}_{n}} \lambda Xy.\; S(X \comp Q)(y) }{ P \trs{\text{exo}_{n}} S }
\rlabel{L$\comp$-exo}
\\[0.5ex]
\rl{ P \comp Q \trs{id} F(A)}{ P \trs{\text{phago}_{n}} F  \quad Q \trs{\overline{\text{phago}}_{n}} A }
\rlabel{id-phago-L}
\quad
\rl{ \cell{\sigma}{P} \trs{id} S(\void)(\sigma')}{ P \trs{\text{exo}_{n}} S \quad \sigma \trs{\coexo_{n}} \sigma' }
\rlabel{id-exo}
\quad
\rl{ \cell{\sigma}{P} \trs{id} \cell{\sigma'}{ \cell{\rho}{\void} \comp P}}{ \sigma \trs{\pino_{n}(\rho)} \sigma' }
\rlabel{id-pino}
\\[0.5ex]
\rl{ \cell{\sigma}{P} \trs{id} \cell{\sigma}{P'} }{ P \trs{id} P' }
\rlabel{id-loc}
\qquad
\rl{ P \comp Q \trs{id} P' \comp Q }{ P \trs{id} P' }
\rlabel{L$\comp$-id}
\end{gather*}
\hrule
\label{tbl:lts}
\caption{Labelled transition system (LTS) for Brane Calculus.
By convention $F : (\sys \to \sys) \to \sys$, $A : \sys \to \sys$ and
$S : \sys \to \mem \to \sys$. Symmetric rules (R$\comp$-phago),
(R$\comp$-$\overline{\text{phago}}$), (R$\comp$-exo), 
(R$\comp$-id), and (id-phago-R) are omitted.}
\end{table}%

An example of how 
labelled transitions are derived using the rules in Table~\ref{tbl:lts}, is
shown in Figure~\ref{fig:derivexample}: the derivation leads to the 
(red-phago) reaction of Table~\ref{tbl:redrules}; note, in particular, 
how the (id-phago) rule is applied.
\begin{figure}[t]
\begin{equation*}
\infr{id-phago-L}{\cell{\phago_n.\sigma | \sigma_0}{P} \comp 
     \cell{\cophago_n(\rho).\tau | \tau_0}{Q} 
     \trs{id}
     \cell{\tau | \tau_0}{\cell{\rho}{\cell{\sigma | \sigma_0}{P}} \comp Q} }{
  \infr{phago}{\cell{\phago_n.\sigma | \sigma_0}{P}
       \trs{\text{phago}_n}
       \lambda Z.\; Z(\cell{\sigma | \sigma_0}{P})}{
    \infr{L-par}{\phago_n.\sigma | \sigma_0 
         \trs{\phago_n}
         \sigma | \sigma_0}{
      \infr{$\phago$-pref}{\phago_n.\sigma \trs{\phago_n} \sigma}{}
      }
    }
  \quad
  \infr{$\overline{\text{phago}}$}{\cell{\cophago_n(\rho).\tau | \tau_0}{Q} 
       \trs{\overline{\text{phago}}_n}
       \lambda X.\; \cell{\tau | \tau_0}{\cell{\rho}{X} \comp Q}
       }{
    \infr{L-par}{\cophago_n(\rho).\tau | \tau_0
        \trs{\cophago_n(\rho)}
        \tau | \tau_0}{
      \infr{$\cophago$-pref}{\cophago_n(\rho).\tau
          \trs{\cophago_n(\rho)}
          \tau}{}
      }
    }
}
\end{equation*}
\caption{Derivation for the (red-phago) reaction.}
\label{fig:derivexample}
\end{figure}
Similar derivations hold for all rules in Table~\ref{tbl:redrules}
except for (red-equiv) (which makes the reduction semantics ``not
structural'').  Hence, it is easy to see that $\trs{id} \subseteq
\react$. The converse follows by an inductive analysis of the
structure of the source processes of an $id$-transition.
\begin{proposition}
  If $P \trs{id} Q$ then $P \react Q$. If $P \react Q$ then $P \trs{id} Q'$ 
  for some $Q' \equiv Q$.
\end{proposition}

Labelled transitions are compatible with structural congruence in the
following sense:
\begin{lemma} \label{lm:trsuptoequiv}
  If $P \trs{\alpha} P'$ and $P \equiv Q$ then there exists $Q'$ such
  that $Q' \equiv P'$ and $Q \trs{\alpha} Q'$.
\end{lemma}

The labelled transition systems, as usual, induces a bisimulation 
relation on terms. Due to the generalization to meta-syntactic 
terms, the canonical bisimulation considers also $\lambda$-terms. 
Since in Brane Calculus the dynamics can happen only at the level of system terms $P \in \P$, we specialize the definition of bisimulation considering only labels in $\set{id, \text{phago}_n, 
\overline{\text{phago}}_n, \text{exo}_n \mid n \in \Lambda }$ as follows:
\begin{definition}[Strong bisimulation]
\label{def:bisim}
  A \emph{bisimulation on Brane Calculus systems} is an 
  equivalence relation $\mathcal{R} \subseteq \P \times \P$ such that 
  for arbitrary $P, Q \in \P$, $(P,Q) \in \mathcal{R}$ if and only if
  \begin{itemize}[label={--}]
    \item if $P \trs{id} P'$ then $\exists\: Q'$ such that 
      $Q \trs{id} Q'$ and $(P',Q') \in \mathcal{R}$;
    \item if $P \trs{\text{phago}} F$ then $\exists\: G$ such that 
      $Q \trs{\text{phago}_n} G$ and $(F(R),G(R)) \in \mathcal{R}$ for all
      $\sigma, \rho: \mem$ and $R' : \sys$, where
      $R = \lambda X.\; \cell{\sigma}{\cell{\rho}{X} \comp R'}$;
    \item if $P \trs{\overline{\text{phago}}} A$ then $\exists\: B$ such that 
      $Q \trs{\overline{\text{phago}}_n} B$ and $(A(R),B(R)) \in \mathcal{R}$ 
      for all $R : \sys$;
    \item if $P \trs{\text{exo}} T$ then $\exists\: S$ such that 
      $Q \trs{\text{exo}_n} S$ and $(T(R)(\rho),S(R)(\rho)) \in \mathcal{R}$ 
      for all $R : \sys$ and $\rho : \mem$;
  \end{itemize}
  Two systems $P,Q \in \P$ are said \emph{bisimilar}, written $P \sim Q$,
  iff there exists a bisimulation relation $\mathcal{R}$ such that 
  $(P,Q)\in \mathcal{R}$.
\end{definition}
In the definition above, when a transition yields a
$\lambda$-abstracted term $M$, this term is instantiated according to
its form. The idea is to recover a bisimulation on only system terms
by sufficiently instantiating the $\lambda$-abstraction in order to
recover a $\sys$-typed term, hence a term in $\P$.  This definition is
slightly different from that in \cite{rs:concur08} for Ambient
Calculus, where Rathke and Sobocinski prefer to add to the LTS
derivation rules for explicit instantiation.  Although this design
choice is equivalent from the point of view of the resulting
bisimulation relation, it leads to an infinitely-branching LTS, since
they chose to endow labels with the instantiation parameters.  In the
definition of the stochastic semantics for the Brane Calculus we need
the LTS to be finitely-branching, hence we do not add instantiation
rules.

The next lemma follows directly from Lemma~\ref{lm:trsuptoequiv}.\begin{lemma}
For arbitrary $P, Q \in \P$, if $P \equiv Q$ then $P \sim Q$.
\end{lemma}
This ensures that the structural equivalence $\equiv$ is 
contained in $\sim$, but we can prove also a stronger result, that is 
${\equiv} \varsubsetneq {\sim}$. 
Let $P = \cell{\zero}{\cell{\phago_n}{}}$; it is easy to see that
$P \sim \void$, however $P \not\equiv \void$, hence structural 
equivalence does not coincide with strong bisimulation, and
in particular $\sim$ equates more terms than $\equiv$. 

\section{A Stochastic Semantics for Brane Calculus}
\label{sec:sbc}

In this section we present a stochastic semantics for the 
Brane calculus, following the construction of \cite{cm:quest10}.
We assume the reader to be familiar with 
basic notions from measure theory (see 
Appendix~\ref{app:measuretheory} for a brief summary of the used definitions).
We start introducing the notation used hereafter and recalling the
definition of \emph{Markov process} (MP) and stochastic bisimulation
on them (more details are in \cite{cm:quest10}).

For arbitrary sets $A$ and $B$, $2^{A}$ denotes the powerset of $A$,
and both $[A \to B]$ and $B^{A}$ will be used to denote the class of
functions from $A$ to $B$. As usual $\N$, $\Q$, and $\R$ denote the
sets of natural, rational, and real numbers, and $\Q^{+}$ and $\R^{+}$
the sets of positive rational and real numbers (with zero),
respectively.  Given a measurable space $(M, \Sigma)$, the elements of
$\Sigma$ are called \emph{measurable sets} and $M$ the
\emph{support-set}. Let $\Delta(M,\Sigma)$ be the class of measures
$\mu \colon \Sigma \to \R^{+}$ on $(M, \Sigma)$.  From
$\Delta(M,\Sigma)$ we distinguish two measures: the \emph{null
  measure} $\omega$ for which $\omega(A) = 0$ for all $A \in \Sigma$,
and, for a fixed $r \in \R^{+}$, the \emph{$r$-Dirac measure on $N$},
$D(r,N)$, defined by $D(r, N)(\bigcup_{i \in I} N_{i}) = \sum_{i \in
  I} f_{N}(N_{i})$, for $N$ and $N_{i \in I}$ elements of a base for
$(M,\Sigma)$ (thus $\bigcup_{i \in I} N_{i} \in \Sigma$), where
$f_{N}(N') = r$ if $N' = N$, otherwise $0$.  Given two measurable
spaces $(M, \Sigma)$ and $(N, \Theta)$, we use $\func{M \to N}$ to
denote the class of measurable functions $f \colon M \to N$ from $(M,
\Sigma)$ to $(N, \Theta)$.

\begin{definition}[Markov kernels and Markov processes]
Let $(M, \Sigma)$ be an measurable space, and $A$ a denumerable set 
of labels. An \emph{$A$-Markov kernel} is a tuple 
$\mathcal{M} = (M, \Sigma, \theta)$, with 
\[
\theta \colon A \to \func{M \to \Delta(M, \Sigma)}.
\]
An \emph{$A$-Markov process} of $\mathcal{M}$ with $m \in M$ as
\emph{initial state}, written $(\mathcal{M},m)$, is the tuple 
$(M, \Sigma, \theta, m)$.
\end{definition}

A MP involves a set $A$ of labels which represent all possible
interactions with the environment.  If $\alpha \in A$ is a label, $m$
is the current state of the system, and $N$ is a measurable set of
states, the function $\theta(\alpha)(m)$ is a measure on the state
space and $\theta(\alpha)(m)(N) \in \R^{+}$ represents the \emph{rate}
of an exponentially distributed random variable characterizing the
duration of the $\alpha$-transition from $m$ to arbitrary $n \in N$.

Given a binary relation $\mathcal{R} \subseteq M \times M$, we call a
subset $N \subseteq M$ $\mathcal{R}$-closed iff
$\mathcal{R} \cap (N \times M) \subseteq N \times N$. 
Given $(M, \Sigma)$ a measurable space and 
$\mathcal{R} \subseteq M \times M$ a binary relation over $M$, 
with $\Sigma(\mathcal{R})$ we denote the set of measurable 
$\mathcal{R}$-closed subsets of $M$.

\begin{definition}[Stochastic bisimulation]
For an $A$-Markov kernel $\mathcal{M}=(M, \Sigma, \theta)$ a
\emph{rate-bisimulation relation} is an equivalence relation 
$\mathcal{R} \subseteq M \times M$ such that $(m,n) \in \mathcal{R}$ iff
for any $C \in \Sigma(\mathcal{R})$ and $\alpha \in A$,
\begin{equation*}
\theta(\alpha)(m)(C) = \theta(\alpha)(n)(C).
\end{equation*}
Two Markov processes $(\mathcal{M}, m)$ and $(\mathcal{M}, n)$ are
\emph{stochastic bisimilar}, written $m \sim_{\mathcal{M}} n$, if $m$ and
$n$ are related by a rate-bisimulation relation.
\end{definition}

In the rest of the section we define the analytic space of terms, and we
show how it can be organized as an $A$-Markov kernel. 
This will implicitly give a stochastic structural operational semantics
such that the canonical behavioral equivalence coincides with the
bisimulation of MPs.

Let us define the measurable space of terms. The construction
takes place at the level of the meta-syntactic terms, hence we will
assume to work with terms in $\T$.
Let $\T/_{\equiv}$ be the set of $\equiv$-equivalence classes on $\T$.
For arbitrary $M \in \T$, we denote by $[M]_{\equiv}$ the 
$\equiv$-equivalence class of $M$. Note that $\T/_{\equiv}$ is a denumerable partition of $\T$, hence it is a generator for a 
$\sigma$-algebra on $\T$. 
\begin{definition}[Measurable space of terms]
The \emph{measurable space of terms} $(\T, \Sigma)$ is the measurable
space on $\T$ where $\Sigma$ is the $\sigma$-algebra on $\T$
generated by $\T/_{\equiv}$.
\end{definition}
The measurable sets are (possibly denumerable) reunions of 
$\equiv$-equivalence classes on $\T$.
In the following we use $\mset{T}$, $\mset{S}$, \dots
to denote arbitrary measurable sets of $\Sigma$.

In order to define an $A$-Markov kernel on $(\T, \Sigma)$ we first
need to define the set of labels.
\begin{definition}[Transition labels]
  The set of \emph{transition labels} for (meta-syntactic) terms is
  given by the pair $(\A, \iota)$, where $\A \triangleq \A_\sys \cup
  \A_\mem$ and
\begin{align*}
  \textsc{(system labels)}  && \A_{\sys} & \triangleq
      \set{\text{phago}_n, 
      \overline{\text{phago}}_n, 
      \text{exo}_n  \mid  n \in \Lambda}
  \\
  \textsc{(membrane labels)} && \A_{\mem} &  \triangleq
      \set{\phago_n, \cophago_n(\rho), 
      \exo_n, \coexo_n, 
      \pino_n(\rho) \mid  n \in \Lambda \text{ and } \rho \in \M}
\end{align*}
\end{definition}
The \emph{internal action} label $id$ is not in $\A$; we extend $\A$
by defining $\A^{+} \triangleq \A \cup \set{id}$.  It is no accident
that the chosen labels are the same of the LTS of
Section~\ref{sec:lts}; indeed, the construction of the Markov kernel
will be guided by the derivation rules listed in Table~\ref{tbl:lts}.

Since we are defining the semantics of a stochastic calculus,
we equip the set of actions $\text{Act} = \set{\phago_{n}, \cophago_{n},
\exo_{n}, \coexo_{n}, \pino_{n} \mid n \in \Lambda}$ with a weight function
$\iota\colon \text{Act} \to \Q^{+} \setminus \set{0}$, which assigns to 
each action the rate of an exponentially distributed random variable that 
characterizes the duration of the transition induced by the execution
of that particular action. The weight function is such that
$\iota(\phago_n) = \iota(\cophago_n)$ and 
$\iota(\exo_n) = \iota(\coexo_n)$; this characterizes the fact that two
cooperating actions have the same execution rate.

Now, we aim to define the function $\theta \colon \A^{+} \to \func{\T
  \to \Delta(\T,\Sigma)}$, which will conclude the construction of an
$\A^{+}$-Markov process for $(\T, \Sigma)$.  To this end, it is useful
to give some operations on measurable sets, which will ease the
exposition of the inductive construction of $\theta$.  For arbitrary
$\mset{T}, \mset{S} \in \Sigma$ and $M \in \T$,
\begin{gather*}
\mset{T} (\mset{S}) \triangleq 
  \bigcup_{\sub{M : t \to t' \in \mset{T} \\[0.1ex] N : t \in \mset{S}}} 
  [M(N)]_{\equiv}
\qquad
\mset{T}_{\; | M} \triangleq \bigcup_{N | M \in \mset{T}} [N]_{\equiv}
\qquad
\mset{T}_{\; \comp M} \triangleq \bigcup_{N \comp M \in \mset{T}} [N]_{\equiv}
\qquad
\mset{T}_{\cell{M}{}} \triangleq \bigcup_{\cell{M}{N} \in \mset{T}} [N]_{\equiv}
\end{gather*}

The next definition constructs the function 
$\theta \colon \A^{+} \to \func{\T \to \Delta(\T,\Sigma)}$ by induction on
the structure of terms. The intuition is that for arbitrary $M \in \T$, $\mset{M}\in \Sigma$ and $\alpha \in \A^{+}$, $\theta(\alpha)(M)(\mset{M})$
represents the total rate of the $\alpha$ actions from $M$ to (elements of)
$\mset{M}$.

\begin{definition} \label{def:meta-theta}
Let $\theta \colon \A^{+} \to \func{\T \to \Delta(\T, \Sigma)}$ be defined 
by induction on the structure of terms:
\begin{description}[leftmargin=-1ex]
  \item[Case $\zero$:] For any $a \in \A^{+}$, let $\theta(a)(\zero) = \omega$.
  
  \item[Case $\alpha.M$:] For any $a \in \A^{+}$ let 
    $\theta(a)(\alpha.M) = D(\iota(\mathop{\mathsf{act}}(\alpha)), [M]_{\equiv})$ 
    if $a = \alpha$, $\theta(a)(\alpha.M) = \omega$ otherwise.

  \item[Case $M | N$:] For any $a \in \A^{+}$ let
    $\theta(a)(M | N)(\mset{T}) = \theta(a)(N)(\mset{T}_{\; | M}) 
    + \theta(a)(M)(\mset{T}_{\; | N})$.
    
  \item[Case $\lambda X. M$:] For any $a \in \A^{+}$, let 
  $\theta(a)(\lambda X.\; M) = \omega$.
  
  \item[Case $\void$:] For any $a \in \A^{+}$, let $\theta(a)(\void) = \omega$.
  
  \item[Case $\cell{M}{N}$:] For any $a \in \A_{\mem}$ let 
  $\theta(a)(\cell{M}{N}) = \omega$. 
  For all other labels in $\A^{+} \setminus \A_{\mem}$ 
  \begin{align*}
    \theta(\text{phago}_n)(\cell{M}{N})(\mset{T}) &= 
      \theta(\phago_n)(M)
      ([\set{ \sigma \in \M \mid 
      \lambda Z:\sys \to \sys.\; Z(\cell{\sigma}{N}) \in \mset{T}}
      ]_{\equiv})
    \\
    \theta(\overline{\text{phago}}_n)(\cell{M}{N})(\mset{T}) &=
      \sum_{\rho\in \M} \theta(\cophago_n(\rho))(M)
      (\set{ \sigma \in \M \mid 
      \lambda X.\; \cell{\sigma}{ \cell{\rho}{X} \comp N} \in \mset{T}}
      /_{\equiv})
    \\
    \theta(\text{exo}_n)(\cell{M}{N})(\mset{T}) &= 
      \theta(\exo_{n})(M)
      (\set{ \sigma \in \M \mid 
      \lambda Xy.\; \cell{\sigma | y}{X} \comp N \in \mset{T}}
      /_{\equiv})
    \\
    \theta(id)(\cell{M}{N})(\mset{T}) & = 
      \theta(id)(N)(\mset{T}_{\cell{M}{}}) + {}
      \\ & \quad
      \sum_{\rho \in \M}^{n \in \Lambda}
      \theta(\pino_{n}(\rho))(M)
      (\set{ \sigma \in \M \mid 
      \cell{\sigma}{ \cell{\rho}{\void} \comp N} \in \mset{T}}
      /_{\equiv}) 
      + {} \\
      & \quad 
      \sum_{\mset{S}([\void]_{\equiv})(\mset{T}') \subseteq \mset{T}}
        \frac{
          \theta(\text{exo}_n)(N)(\mset{S}) \cdot
          \theta(\cophago_n)(M)(\mset{T}')
          }{\iota(\coexo_n)}
  \end{align*}
  
  \item[Case $M \comp N$:] For any $a \in \A_{\mem}$ let 
  $\theta(a)(M \comp N) = \omega$.
  For all other labels in $\A^{+} \setminus \A_{\mem}$
  \begin{align*}
    \theta(\text{phago}_n)(M \comp N)(\mset{T}) &= 
       \theta(\text{phago}_n)(N)
       (\set{F : (\sys \to \sys) \to \sys \mid 
       \lambda Z.\; (F(Z) \comp M) \in \mset{T})}/_{\equiv}) + {}
       \\ & \phantom{{} = {}}
       \theta(\text{phago}_n)(M)
       (\set{F : (\sys \to \sys) \to \sys \mid 
       \lambda Z.\; (F(Z) \comp N) \in \mset{T})}/_{\equiv})
    \\
    \theta(\overline{\text{phago}}_n)(M \comp N)(\mset{T}) &= 
       \theta(\overline{\text{phago}}_n)(N)
       (\set{A : \sys \to \sys \mid 
       \lambda X.\; (A(X) \comp M) \in \mset{T})}/_{\equiv}) + {}
       \\ & \phantom{{} = {}}
       \theta(\overline{\text{phago}}_n)(M)
       (\set{A : \sys \to \sys \mid 
       \lambda X.\; (A(X) \comp N) \in \mset{T})}/_{\equiv})
    \\
    \theta(\text{exo}_n)(M \comp N)(\mset{T}) &= 
       \theta(\text{exo}_n)(N)
       (\set{T : \sys \to \mem \to \sys \mid
       \lambda Xy.\; T(X \comp M)(y) \in \mset{T}}/_{\equiv}) + {}
       \\ & \phantom{{} = {}}
       \theta(\text{exo}_n)(M)
       (\set{T : \sys \to \mem \to \sys \mid
       \lambda Xy.\; T(X \comp N)(y) \in \mset{T}}/_{\equiv})
    \\
    \theta(id)(M \comp N)(\mset{T}) &= 
      \theta(id)(N)(\mset{T}_{\; \comp M}) +
      \theta(id)(M)(\mset{T}_{\; \comp N}) + {} \\
      &\quad
       \sum^{n \in \Lambda}_{\mset{F}(\mset{A}) \subseteq \mset{T}}
         \frac{ 
           \theta(\text{phago}_n)(M)(\mset{F}) \cdot
           \theta(\overline{\text{phago}}_n)(N)(\mset{A})
         }{\iota(\phago_n)}
         + {} \\
       &\qquad
       \sum^{n \in \Lambda}_{\mset{F}(\mset{A}) \subseteq \mset{T}}
         \frac{ 
           \theta(\text{phago}_n)(N)(\mset{F}) \cdot
           \theta(\overline{\text{phago}}_n)(M)(\mset{A})
         }{\iota(\phago_n)}
  \end{align*}
\end{description}
\end{definition}
The intuition behind this definition is that each summand 
corresponds to a derivation rule of the LTS of Section~\ref{sec:lts}.
For example, in $\theta(id)(M \comp N)(\mset{T})$ the last 
summand corresponds to (id-phago-R) rule. 
Similarly, if there are no $a$-transitions for a term 
$M$ in the LTS, $\theta(a)(M) = \omega$; this is the case of 
$\zero$, $\void$, and $\lambda X.\; M$.
Note that for each $\sigma \in \M$, $\theta(a)(\sigma) \neq \omega$
iff $a \in \text{act}(\sigma)$. Consequently each infinitary sum involved
in Definition~\ref{def:meta-theta} has a finite number of non-zero
summands.

In particular we have a correspondence between the LTS and the
function $\theta$ in the following sense.
\begin{proposition} \label{prop:lstthetacorrespondence}
For arbitrary $M \in T$ and $\alpha \in \A^{+}_{\sys}$ the following statements hold
\begin{enumerate}
  \item if $\theta(\alpha)(M)(\mset{T}) > 0$ then there exists 
   $M' \in \mset{T}$ such that $M \trs{\alpha} M'$,
  \item if $M \trs{\alpha} M'$ then there exists $\mset{M} \in \Pi$ 
  such that $M' \in \mset{T}$ and $\theta(\alpha)(M)(\mset{T}) > 0$.
\end{enumerate}
\end{proposition}
In the proposition above, (1) can be proven by induction on the
structure of the term $M$ (assumed to be well-typed); while 
the prove for (2) is by induction on the derivation of $M \trs{\alpha} M'$.
Note that Proposition~\ref{prop:lstthetacorrespondence} reflects the
similarity between Definition~\ref{def:meta-theta} and the definition
of the LTS of Section~\ref{sec:lts}.

A direct consequence of Proposition~\ref{prop:lstthetacorrespondence} 
is the following.
\begin{corollary} \label{cor:lstthetacorrespondence}
$M \trs{\alpha} M'$ iff $\theta(\alpha)(M)([M']_{\equiv}) > 0$.
\end{corollary}

The next theorem states that $(\T, \Sigma, \theta)$ is an 
$\A^{+}$-Markov kernel. Proving this will implicitly show the correctness
of our construction, indeed it suffices to prove that for each $M \in \T$
and each $\alpha \in \A^{+}$, $\theta(\alpha)(P) : \Sigma \to \R^{+}$
is a measure on the measurable space $(\T, \Sigma)$.
\begin{theorem}[Markov kernel for the measurable space of terms]
$(\T, \Sigma, \theta)$ is an $\A^{+}$-Markov kernel.
\end{theorem}
A consequence of this theorem is that for each $M \in \T$, 
$(\T, \Sigma, \theta, M)$ is a Markov process, hence we can define
a stochastic bisimulation for Brane Calculus meta-syntactic terms,
simply as the stochastic bisimulation of Markov processes in 
$(\T, \Sigma, \theta)$.

From the measurable space $(\T, \Sigma)$ we can recover 
the measurable subspaces $(\M, \Theta)$ and $(\P, \Pi)$, respectively
for membrane and system terms. Indeed, if $\Theta$ is the 
$\sigma$-algebra generated from the base $\M/_{\equiv}$, and 
$\Pi$ the one generated from the base $\P/_{\equiv}$, both $\Theta$
and $\Pi$ are contained in $\Sigma$, hence $(\M, \Theta)$ and $(\P, \Pi)$
are two subspaces of $(\T, \Sigma)$.

One may be led to think that by a suitable restriction on $\theta$
we can also recover two $A$-Markov kernels respectively on the 
sets $\A_{\mem}$ and $\A_{\sys}^{+}$ of labels.
Of course, it is possible to define two Markov kernels by letting
$\theta_{\mem} \colon \A_{\mem}^{+} \to \func{\M \to \Delta(\M,\Theta)}$
and $\theta'_{\sys} \colon \A_{\sys}^{+} \to \func{\P \to \Delta(\P,\Pi)}$ as
\begin{align*}
\theta_{\mem}(a)(\sigma)(\mset{M}) &= \theta(a)(\sigma)(\mset{M})
&& \text{for all $\mset{M} \in \Theta$}
\\
\theta'_{\sys}(b)(P)(\mset{P}) &= \theta(b)(P)(\mset{P})
&& \text{for all $\mset{P} \in \Pi$}.
\end{align*}
\noindent
Although this definition works well for $(\M, \Theta)$, it does not
straightforwardly work for $(\P, \Pi)$. In fact $\theta'_{\sys}$ does not enjoy
a result similar to Proposition~\ref{prop:lstthetacorrespondence}, and
moreover the stochastic bisimulation is not the equivalence that one may
expect.
For example, it is easy to see that for all $\alpha \in \A_{\sys}^{+}$
and $\mset{P} \in \Pi$, $\theta(\alpha)(\cell{\exo_{n}}{\void})(\mset{P}) = 0$.
The same thing happen for $\cell{\phago_{m}}{\void}$, that is, 
$\theta(\alpha)(\cell{\phago_{n}}{\void})(\mset{P}) = 0$, hence 
$(\P, \Pi, \theta'_{\sys}, \cell{\exo_{n}}{\void})$ and 
$(\P, \Pi, \theta'_{\sys}, \cell{\phago_{m}}{\void})$ are stochastic 
bisimilar. 
By a simple analysis on the structure of $\cell{\exo_n}{\void}$ and 
$\cell{\phago_m}{\void}$, it is easy to see that 
\begin{gather*}
\theta(\text{exo}_{n})(\cell{\exo_n}{\void})
([\lambda Xy.\; \cell{\zero | y}{X} \comp \void)]_{\equiv})
=  
\theta(\text{exo}_{n})(\cell{\exo_n}{\void})
([\lambda Xy.\; \cell{\zero | y}{X})]_{\equiv})
= 
\iota(\exo_{n}) > 0
\\[1ex]
\theta(\text{phago}_{m})(\cell{\phago_m}{\void})
([\lambda Z.\; Z(\cell{\zero}{\void})]_{\equiv})
=  
\theta(\text{phago}_{m})(\cell{\phago_m}{\void})
([\lambda Z.\; Z(\void)]_{\equiv}) 
= 
\iota(\phago_{m}) > 0.
\end{gather*}
We can prove that 
$\theta(\text{exo}_{n})(\cell{\exo_n}{\void})(\mset{T}) > 0$ iff 
$\lambda Xy.\; \cell{\zero | y}{X} \in \mset{T}$, and 
$\theta(\text{phago}_{m})(\cell{\phago_m}{\void})(\mset{T}) > 0$ iff
$\lambda Z.\; Z(\void) \in \mset{T}$ (remember that 
both $\theta(\text{exo}_{n})(\cell{\exo_n}{\void})$ and 
$\theta(\text{phago}_{m})(\cell{\phago_m}{\void})$ are measures).
This example points out the problem: the two measurable
spaces ``are not of the right type''. 
Indeed, in Definintion~\ref{def:meta-theta} it easy to see that
for labels in $\A_{\sys}$ (see cases $\cell{M}{N}$ and $M \comp N$)
$\theta$ contributes to the rate with a nonzero value only if the
the measurable space given as (last) parameter contains
terms of nonbase type. In particular, for label $\text{phago}_{n}$
the type is $(\sys \to \sys) \to \sys$, for label $\overline{\text{phago}}_{n}$
is $\sys \to \sys$, and for label $\text{exo}_{n}$ is $\sys \to \mem \to \sys$.

This simple consideration induces the definition of another function:
\begin{definition}
Let $\tilde{\theta}_{\sys} \colon \A_{\sys}^{+} \to \func{\P \to \Delta(\P, \Pi)}$ 
be defined as follows, where $P \in \P$ and $\mset{P} \in \Pi$.
\begin{align*}
\tilde{\theta}_{\sys}(id)(P)(\mset{P}) &= \theta(id)(P)(\mset{P})
\\[0.5ex]
\tilde{\theta}_{\sys}(\text{phago}_{n})(P)(\mset{P}) &= 
  \theta(\text{phago}_{n})(P)(\mset{F}) 
  \\
  & \qquad \text{where }  
  \mset{F} = \set{ F : (\sys \to \sys) \to \sys 
  \mid \exists\: A: \sys \to \sys. \; F(A) \in \mset{P}}/_{\equiv}
\\[0.5ex]
\tilde{\theta}_{\sys}(\overline{\text{phago}}_{n})(P)(\mset{P}) &=
  \theta(\overline{\text{phago}}_{n})(P)(\mset{A})
  \\
  & \qquad \text{where }  
  \mset{A} = \set{ A: \sys \to \sys 
  \mid \exists\: F : (\sys \to \sys) \to \sys. \; F(A) \in \mset{P}}/_{\equiv}
\\[0.5ex]
\tilde{\theta}_{\sys}(\text{exo}_{n})(P)(\mset{P}) &=
  \theta(\text{exo}_{n})(P)(\mset{S})
  \\
  & \qquad \text{where }  
  \mset{S} = \set{ S: \sys \to \mem \to \sys 
  \mid \exists\: Q \in \P, \sigma \in \M. \; S(Q)(\sigma) \in \mset{P}}/_{\equiv}
\end{align*}
\end{definition}
Again $(\P, \Sigma, \tilde\theta_{\sys})$ is an $\A_{\sys}^{+}$-Markov
kernel, and in particular the stochastic bisimulation on Markov
processes is the one that we need.  It is easy to see that $(\P, \Pi,
\tilde\theta_{\sys}, \cell{\exo_{n}}{\void})$ and $(\P, \Pi,
\tilde\theta_{\sys}, \cell{\phago_{m}}{\void})$ are no more stochastic
bisimilar (assuming $\iota(\exo)_{n} \neq \iota(\phago)_{n}$).

However, the above definition is not defined inductively on the
structure of terms. To this end, we provide another function
$\theta_{\sys}\colon \A^{+}_{\sys} \to \func{\P \to \Delta(\P, \Pi)}$,
defined by induction on the structure of system terms, and we prove
that $\theta_{\sys}$ is an alternative characterization of
$\tilde\theta_{\sys}$, i.e., $\theta_{\sys}$ and $\tilde\theta_{\sys}$
coincide.

\begin{definition} \label{def:thetasystems}
Let $\theta_{\sys}\colon \A^{+}_{\sys} \to \func{\P \to \Delta(\P, \Pi)}$
be defined on the structure of $P \in \P$, as follows.
\begin{description}[leftmargin=-1ex]
  \item[Case $P = \void$:]
  For any $\alpha \in \A^{+}_{\sys}$, let $\theta_{\sys}(\alpha) = \omega$.
  
  \item[Case $P = \cell{\sigma}{Q}$:]
  For any $\mset{P} \in \Pi$,
  \begin{align*}
  \theta_{\sys}(\text{phago}_n)(\cell{\sigma}{Q})(\mset{P}) &=
    \theta_{\mem}(\phago_n)(\sigma)
    (\set{ \sigma' \mid 
    \cell{\tau}{ \cell{\rho}{\cell{\sigma'}{Q}} \comp R} \in \mset{P}
    }/_{\equiv})
  \\
  \theta_{\sys}(\overline{\text{phago}}_n)(\cell{\sigma}{Q})(\mset{P}) &=
    \sum_{\rho \in \M}
    \theta_{\mem}(\cophago_n(\rho))(\sigma)
    (\set{ \sigma' \mid 
    \cell{\sigma'}{ \cell{\rho}{\cell{\tau}{R}} \comp Q} \in \mset{P}
    }/_{\equiv})
  \\
  \theta_{\sys}(\text{exo}_n)(\cell{\sigma}{Q})(\mset{P}) &=
    \theta_{\mem}(\exo_n)(\sigma)
    (\set{ \sigma' \mid 
    \cell{\sigma' | \tau}{R} \comp Q \in \mset{P}
    }/_{\equiv})
  \\
  \theta_{\sys}(id)(\cell{\sigma}{Q})(\mset{P}) &=
    \theta_{\sys}(id)(Q)(\mset{P}_{\cell{\sigma}{}}) + {} \\
    & \quad \sum_{\rho \in \M}^{n \in \Lambda}
    \theta_{\mem}(\pino_n(\rho))(\sigma)
    (\set{\sigma' \mid
    \cell{\sigma'}{\cell{\rho}{\void} \comp Q} \in \mset{P}}/_{\equiv})
    + {}
    \\
    & \quad \sum_{n \in \Lambda}
      \frac{
      \theta_{\mem}(\coexo_{n})(\sigma)
      (\set{ \sigma' \mid \cell{\sigma' | \tau}{Q} \comp R \in \mset{P}}/_{\equiv})
      \cdot
      \theta_{\sys}(\text{exo}_n)(Q)(\mset{P})
      }{\iota(\phago_n)}
  \end{align*}
  
  \item[Case $P = Q \comp R$:]
  For all $\alpha \in \A_{\sys}$ and $\mset{P} \in \Pi$
  \begin{align*}
  \theta_{\sys}(\alpha)(Q \comp R)(\mset{P}) &=
    \theta_{\sys}(\alpha)(R)(\mset{P}_{\; \comp Q}) +
    \theta_{\sys}(\alpha)(Q)(\mset{P}_{\; \comp R})
  \\
  \theta_{\sys}(id)(Q \comp R)(\mset{P}) &=
    \theta_{\sys}(id)(R)(\mset{P}_{\; \comp Q}) +
    \theta_{\sys}(id)(Q)(\mset{P}_{\; \comp R}) + {} \\
    & \quad \sum_{n \in \Lambda}
    \frac{
     \theta_{\sys}(\text{phago}_n)(R)(\mset{P}) \cdot
     \theta_{\sys}(\overline{\text{phago}}_n)(Q)(\mset{P})
    }{\iota(\phago_n)} + {} \\
    & \qquad \sum_{n \in \Lambda}
    \frac{
     \theta_{\sys}(\text{phago}_n)(Q)(\mset{P}) \cdot
     \theta_{\sys}(\overline{\text{phago}}_n)(R)(\mset{P})
    }{\iota(\phago_n)}
  \end{align*}
\end{description}
\end{definition}

The next theorem states that $\theta_{\sys}$ precisely characterizes
the function $\tilde\theta_{\sys}$ defined before.
\begin{theorem}
For all $\alpha \in \A^{+}_{\sys}$, $P \in \P$, and $\mset{P} \in \Pi$,
$\theta_{\sys}(\alpha)(P)(\mset{P}) = 
\tilde\theta_{\sys}(\alpha)(P)(\mset{P})$.
\end{theorem}
A consequence of the previous theorem is that $(\P, \Pi, \theta_{\sys})$
is an $\A^{+}_{\sys}$-Markov kernel, and that the stochastic
bisimulation on Markov processes coincides with that induced by
$(\P, \Pi, \tilde\theta_{\sys})$.

\section{Stochastic Structural Operational Semantics and Bisimulation}\label{sec:bisim}

In this section we introduce the stochastic structural operational
semantics for the Brane Calculus, with the aim of defining a
behavioral equivalence on system terms that coincides with their
bisimulation as Markov processes on $(\P, \Pi, \theta_{\sys})$.
Notably, it is directly induced from the definition of the function
$\theta_{\sys}$ (Definition~\ref{def:thetasystems}), following the
pattern of \cite{cm:quest10}.
Since this is an unusual construction for a structural operational
semantics, some preliminary discussions are needed.
Typically, a structural operational semantics for a stochastic process algebra associates a rate $r$ with transitions: $P \trs{\alpha}_{r} P'$.
Instead, in  \cite{cm:quest10} transitions are not between two processes,
but from a process to an infinite measurable set of processes.
In order to maintain ``the spirit'' of process algebras Cardelli and Mardare
replace the classic rules of the form $P \trs{\alpha} P'$ with rules of the form 
$P \to \mu$, where $\mu$ is a function from action labels to measures on 
the measurable space of processes.
Let us see how this construction can be applied to the Brane Calculus.

For simplifying the rules of the operational semantics, we
first define some operations on the functions in 
$\Delta(\M, \Theta)^{\A_{\mem}}$ and $\Delta(\P, \Pi)^{\A^{+}_{\sys}}$,
and analyze their properties. 
We say that a function $\mu \in \Delta(\P, \Pi)^{\A^{+}_{\sys}}$ has
\emph{finite support} if $\A^{+}_{\sys} \setminus \mu^{-1}(\omega)$ is
finite or empty (recall that $\omega$ is the null measure).

\begin{definition}
\label{def:measop}
Consider the following constants and operations on 
$\Delta(\M, \Theta)^{\A_{\mem}}$ defined as follows.
\begin{itemize}
  \item $\omega^{\mem} \colon \A_{\mem} \to \Delta(\M, \Theta)$
  defined as $\omega^{\mem}(\alpha) = \omega$, for arbitrary 
  $\alpha \in \A_{\mem}$; 

  \item For arbitrary $a \in \A_{\mem}$ and $\sigma, \rho \in \M$, let
  $\alpha_{\sigma} \colon \A_{\mem} \to \Delta(\M, \Theta)$ be defined by
  \begin{align*}
   [\epsilon]_{\sigma}(a) &= 
    \begin{cases}
      D(\iota(\epsilon), [\sigma]_{\equiv}) & \text{if $a = \epsilon$} \\
      \omega & \text{if $a \neq \epsilon$}
    \end{cases}
    & \text{(for 
    $\epsilon \in \set{\phago_{n}, \exo_{n}, \coexo_{n} \mid n \in \Lambda}$)}
    \\
    [\epsilon(\rho)]_{\sigma}(a) &= 
    \begin{cases}
      D(\iota(\epsilon), [\sigma]_{\equiv}) & \text{if $a = \epsilon(\rho)$} \\
      \omega & \text{if $a \neq \epsilon(\rho)$}
    \end{cases}
    & \text{(for 
    $\epsilon \in \set{\cophago_{n}, \pino_{n} \mid n \in \Lambda}$)}
  \end{align*}
  
  \item For arbitrary $\mu, \mu' \in \Delta(\M, \Theta)^{\A_{\mem}}$
  with finite support, $a \in \A_{\mem}$, $\sigma, \tau \in \M$, and
  $\mset{M} \in \Theta$, let the function 
  $\mu \mpar{\sigma}{\tau} \mu' \A_{\mem} \to \Delta(\M, \Theta)$ 
  be defined by
  \begin{equation*}
    (\mu \mpar{\sigma}{\tau} \mu')(a)(\mset{M}) =
       \mu(a)(\mset{M}_{\tau}) + \mu'(a)(\mset{M}_{\sigma})
  \end{equation*}
\end{itemize}
Consider the following constants and operations on $\Delta(\P, \Pi)^{\A^{+}_{\sys}}$ and $\Delta(\M, \Theta)^{\A_{\mem}}$
defined as follows.
\begin{itemize}
  \item $\omega^{\sys} \colon \A^{+}_{\sys} \to \Delta(\P, \Pi)$
  defined as $\omega^{\sys}(\alpha) = \omega$, for arbitrary 
  $\alpha \in \A^{+}_{\sys}$;
  
  \item For arbitrary $\mu \in \Delta(\P, \Pi)^{\A^{+}_{\sys}}$, 
  $\mu' \in \Delta(\M, \Theta)^{\A_{mem}}$ with finite support, 
  $\sigma \in \M$, $P \in \P$, and $\mset{P} \in \Pi$ let the function 
  $\mu \mAt{\sigma}{P} \mu' \colon \A^{+}_{\sys} \to \Delta(\P, \Pi)$
  be defined by:
  \begin{align*}
    (\mu \mAt{\sigma}{P} \mu')(\text{phago}_n)(\mset{P}) &=
      \mu'(\phago_n)
      (\set{ \sigma' \mid 
      \cell{\tau}{ \cell{\rho}{\cell{\sigma'}{P}} \comp Q} \in \mset{P}
      }/_{\equiv})
    \\
    (\mu \mAt{\sigma}{P} \mu')(\overline{\text{phago}}_n)(\mset{P}) &=
      \sum_{\rho \in \M}
      \mu'(\cophago_n(\rho))
      (\set{ \sigma' \mid 
      \cell{\sigma'}{ \cell{\rho}{\cell{\tau}{Q}} \comp P} \in \mset{P}
      }/_{\equiv})
    \\
    (\mu \mAt{\sigma}{P} \mu')(\text{exo}_n)(\mset{P}) &=
      \mu'(\exo_n)
      (\set{ \sigma' \mid 
      \cell{\sigma' | \tau}{Q} \comp P \in \mset{P}
      }/_{\equiv})
    \\
    (\mu \mAt{\sigma}{P} \mu')(id)(\mset{P}) &=
      \mu(id)(\mset{P}_{\cell{\sigma}{}}) + {}
      \sum_{\rho \in \M}^{n \in \Lambda}
      \mu'(\pino_n(\rho))
      (\set{\sigma' \mid
      \cell{\sigma'}{\cell{\rho}{\void} \comp Q} \in \mset{P}}/_{\equiv}) + {}
      \\ &
      \sum_{n \in \Lambda}
      \frac{ 
      \mu'(\coexo_n)
      (\set{ \sigma' \mid \cell{\sigma' | \tau}{P} \comp Q \in \mset{P}}/_{\equiv})
      \cdot
      \mu(\text{exo}_n)(\mset{P})
      }{\iota(\phago_n)}
  \end{align*}
  
  \item For arbitrary $\mu, \mu' \in \Delta(\P, \Pi)^{\A^{+}_{\sys}}$
  with finite support, $a \in \A_{\sys}$, $P, Q \in \P$, and 
  $\mset{P} \in \Pi$, 
  let the function 
  $\mu \mtimes{P}{Q} \mu' \colon \A^{+}_{\sys} \to \Delta(\P, \Pi)$ 
  be defined by
  \begin{align*}
    (\mu \mtimes{P}{Q} \mu')(a)(\mset{P}) &=
       \mu(a)(\mset{P}_{Q}) + \mu'(a)(\mset{P}_{P})
    \\
    (\mu \mtimes{P}{Q} \mu')(id)(\mset{P}) &=
      \mu(id)(\mset{P}_{Q}) + \mu'(id)(\mset{P}_{P}) + {}
      \\ &
      \sum_{n \in \Lambda}
      \frac{
      \big(\mu(\text{phago}_n)(\mset{P}) \cdot 
      \mu'(\overline{\text{phago}}_n)(\mset{P})\big)
      +
      \big( \mu(\overline{\text{phago}}_n)(\mset{P}) \cdot
      \mu'(\text{phago}_n)(\mset{P})\big)
      }{\iota(\phago_n)}
  \end{align*}
\end{itemize}

\end{definition}
Notice that since $\mu$ and $\mu'$ have finite support, each
infinite sum involved in Definition~\ref{def:measop} has a finite number
of nonzero summands.
The next two lemmata prove that the definitions of $\mpar{\sigma}{\tau}$,
$\mtimes{P}{Q}$, and $\mAt{\sigma}{P}$, for arbitrary $\sigma, \tau \in \M$
and $P, Q \in \P$ are correct; they also state some basic properties
of these operators.
\begin{lemma} The following statements hold.
\begin{enumerate}
 \item For arbitrary $\sigma, \tau, \rho \in \M$ and 
 $\mu', \mu'', \mu''' \in \Delta(\M, \Theta)^{\A_{\mem}}$ with finite support
 \begin{enumerate}
   \item $\mu' \mpar{\sigma}{\tau} \mu'' = \mu'' \mpar{\tau}{\sigma} \mu'$,
   \item $(\mu' \mpar{\sigma}{\tau} \mu'') \mpar{\sigma|\tau}{\rho} \mu''' =
     \mu' \mpar{\sigma}{\tau|\rho} (\mu'' \mpar{\tau}{\rho} \mu''')$,
   \item $\mu' \mpar{\sigma}{\, \zero} \omega^{\mem} = \mu'$.
 \end{enumerate}
 
 \item For arbitrary $P, Q, R \in \P$ and 
 $\mu', \mu'', \mu''' \in \Delta(\P, \Pi)^{\A^{+}_{\sys}}$ with finite support
 \begin{enumerate}
   \item $\mu' \mtimes{P}{Q} \mu'' = \mu'' \mtimes{Q}{P} \mu'$,
   \item $(\mu' \mtimes{P}{Q} \mu'') \mtimes{P|Q}{R} \mu''' =
     \mu' \mtimes{P}{Q|R} (\mu'' \mtimes{Q}{R} \mu''')$,
   \item $\mu' \mtimes{P}{\void} \omega^{\sys} = \mu'$.
 \end{enumerate}
 \item $\omega^{\sys} \mAt{\zero}{\void} \omega^{\mem} = \omega^{\sys}$.
\end{enumerate}
\end{lemma}

\begin{lemma} The following statements hold.
\begin{enumerate}
 \item For arbitrary $\sigma, \sigma', \tau, \tau' \in \M$ and 
 $\mu', \mu'' \colon \A_{\mem} \to \Delta(\M, \Theta)$ with finite support
   \begin{enumerate}
     \item for $\epsilon \in \set{\phago_{n}, \exo_{n}, \coexo_{n} \mid 
     n \in \Lambda}$, 
     $\sigma \equiv \tau$ implies $[\epsilon]_{\sigma} = [\epsilon]_{\tau}$,
     
     \item for $\epsilon \in \set{\cophago_{n}, \pino_{n} \mid 
     n \in \Lambda}$, 
     $\sigma \equiv \tau$ and $\rho \equiv \rho'$ imply 
     $[\epsilon(\rho)]_{\sigma} = [\epsilon(\rho')]_{\tau}$,
     
     \item $\sigma \equiv \sigma'$ and $\tau \equiv \tau'$ imply 
       $\mu' \mpar{\sigma}{\tau} \mu'' = \mu' \mpar{\sigma'}{\tau'} \mu''$.
   \end{enumerate}
   
 \item For arbitrary $P, P', Q, Q' \in \P$, $\sigma, \tau \in \M$, 
 $\mu, \mu', \mu'' \colon \A^{+}_{\sys} \to \Delta(\P, \Pi)$, and
 $\nu \colon \A_{\mem} \to \Delta(\M, \Theta)$ with finite support
   \begin{enumerate}
     \item $P \equiv Q$ and $\sigma \equiv \tau$ imply
       $\mu \mAt{\sigma}{P} \nu = \mu \mAt{\tau}{Q} \nu$,
       
     \item $P \equiv P'$ and $Q \equiv Q'$ imply
       $\mu' \mtimes{P}{Q} \mu'' = \mu' \mtimes{P'}{Q'} \mu''$.
   \end{enumerate}
\end{enumerate}
\end{lemma}

The rules of the operational semantics are listed in 
Table~\ref{tbl:stochSOS}. 
\begin{table}[t]
\hrule
\begin{gather*}
\rl{\zero \to \omega^{\mem}}{}
\rlabel{zero}
\quad
\rl{\epsilon.\sigma \to [\epsilon]_{\sigma}}{
  \epsilon \in \set{\phago_n, \exo_n, \coexo_n}}
\rlabel{pref}
\quad
\rl{\epsilon(\rho).\sigma \to [\epsilon(\rho)]_{\sigma}}{
  \epsilon \in \set{\cophago_n, \pino_n}}
\rlabel{pref-arg}
\quad
\rl{\sigma | \tau \to \mu' \mpar{\sigma}{\tau} \mu''}{
\sigma \to \mu' \qquad \tau \to \mu''}
\rlabel{par}
\\[2ex]
\rl{ \void \to \omega^{\sys}}{}
\rlabel{void}
\qquad\qquad
\rl{ \cell{\sigma}{P} \to \mu \mAt{\sigma}{P} \nu}{
  \sigma \to \nu \qquad P \to \mu}
\rlabel{loc}
\qquad\qquad
\rl{ P \comp Q \to \mu' \mtimes{P}{Q} \mu''}{
  P \to \mu' \qquad Q \to \mu''}
\rlabel{comp}
\end{gather*}
\hrule
\caption{Structural operational semantics for Brane Calculus}
\label{tbl:stochSOS}
\end{table}
The operational semantics associates with
each membrane $\sigma$ a mapping $\nu \in \Delta(\M, \Theta)^{\A_{mem}}$, and with each system $P \in \P$ a mapping
$\mu \in \Delta(\P, \Pi)^{\A^{+}_{\sys}}$. 
For each $\equiv$-closed set $\mset{M} \in \Theta$ and each label
$\alpha \in \A_{mem}$, $\nu(\alpha)(\mset{M}) \in \R^{+}$ represents
the total rate of an $\alpha$-transition of $\sigma$ to some arbitrary
element in $\mset{M}$; and for each set $\mset{P} \in \Pi$ and label
$\alpha' \in \A^{+}_{\sys}$, similarly $\mu(\alpha')(\mset{P}) \in \R^{+}$ represents
the total rate of an $\alpha'$-transition of $P$ to some arbitrary
element in $\mset{P}$.

The next lemma guarantees the consistency of the stochastic transition relation $\to$, and as a consequence the consistency of the operational
semantics.
\begin{lemma}[Uniqueness of the measure]
\label{lm:uniqueness} For each $\sigma\in \M$ and $P\in \P$:
\begin{enumerate}
  \item there exists a unique 
    $\mu \in \Delta(\M,\Theta)^{\A_{\mem}}$ such that $\sigma \to \mu$;
    moreover, $\mu$ has finite support;
  \item there exists a unique 
    $\mu \in \Delta(\P,\Pi)^{\A_{\sys}}$ such that $P \to \mu$;
    moreover, $\mu$ has finite support.
\end{enumerate}
\end{lemma}

This operational semantics can be further used to define various 
pointwise semantics as, e.g.:
\begin{equation*}
  P \trs{\alpha, r} Q 
  \qquad \text{iff} \qquad  
  P \to \mu \; \text{ and } \; \mu(\alpha)([Q]_{\equiv}) = r
\end{equation*}

\begin{example} \label{ex:inducedpointwise}
Suppose $\iota(\phago_{n}) = r$, and $\tau, \rho \in \M$,  
$Q \in \P$; it is easy to see that
\begin{enumerate}
  \item $\cell{\phago_{n}.\sigma}{P} \trs{\text{phago}_{n}, r} 
              \cell{\tau}{ \cell{\rho}{ \cell{\sigma}{P} } \comp Q}$
  \item $\cell{\phago_{n}.\sigma}{P} \comp \cell{\phago_{n}.\sigma}{P} 
              \trs{\text{phago}_{n}, 2r} 
              \cell{\tau}{ \cell{\rho}{ \cell{\sigma}{P} } \comp Q} \comp 
              \cell{\phago_{n}.\sigma}{P} $
  \item $\cell{\phago_{n}.\sigma}{P} \comp \cell{\phago_{n}}{\void} 
              \trs{\text{phago}_{n}, r} 
              \cell{\tau}{ \cell{\rho}{ \cell{\sigma}{P} } \comp Q} \comp 
              \cell{\phago_{n}}{\void} $
  \item $\cell{\phago_{n}.\sigma}{P} \comp \cell{\phago_{n}}{\void} 
              \trs{\text{phago}_{n}, r} 
              \cell{\phago_{n}.\sigma}{P} \comp
              \cell{\tau}{ \cell{\rho}{ \void } \comp Q}$
  \item $\cell{\phago_{n}.\sigma}{P} \comp \cell{\cophago_{n}(\rho).\tau}{Q} 
              \trs{id, r} 
              \cell{\tau}{ \cell{\rho}{ \cell{\sigma}{P} } \comp Q}$
\end{enumerate}
\end{example}
Note that in (1) the chosen $\tau$, $\rho$, and $Q$ are not relevant,
indeed the ``same'' transition holds also for different choices of them.
At first sight (1) seems curious, but the
intuition behind it is that it is not important where a cell goes once it
is phagocytized, but that eventually it could be phagocytized. 
Examples~\ref{ex:inducedpointwise} (2 -- 4) show that the total rate is correctly summed up,
while (5) exhibits an internal action and how it influences the total rate. 

The next lemma states that operational semantics does not distinguish
structurally equivalent terms:
\begin{lemma} \label{lm:stransitionisuptoequiv}
 Stochastic transitions are up-to structural equivalence:
  \begin{enumerate}
     \item if $\sigma \equiv \tau$ and $\sigma \to \mu$ then $\tau \to \mu$,
     \item if $P \equiv Q$ and $P \to \mu$ then $Q \to \mu$.
  \end{enumerate}
\end{lemma}
Notice that the converse does not hold in general, that is, if for some
$P, Q \in \P$, $P \to \mu'$ and $Q \to \mu''$, and $\mu' = \mu''$, this 
does not imply that $P \equiv Q$. For example, let
$P = \cell{\zero}{\cell{\phago_{n}}{}}$ and 
$Q = \void$, then 
\begin{align*}
P & \to \big( \mu_{1} = 
(\omega_{\sys} \mAt{\phago_{n}}{\void} [\phago_{n}]_{\zero}) 
\mAt{\zero}{\cell{\phago_{n}}{}} \omega^{\mem} 
\big)
\\
Q &\to \big( \mu_{2} = \omega^{\sys} \big)
\end{align*}
It is trivial to verify that $\mu_{1} = \mu_{2}$, however 
$P \not\equiv Q$. In fact for each 
$\alpha \in \A_{\sys}$, $\mu_{1}(\alpha)(\mset{P}) = 0$ (for all
$\mset{P}$), since $\mu_{1}$ is of the form $\mu'_{1} \mAt{\rho}{R} \omega^{\mem}$ and its result depends only on $\omega^{\mem}$.
In order to prove $\mu_{1}(id)(\mset{P}) = 0$ it suffices to
verify that 
$(\omega_{\sys} \mAt{\phago_{n}}{\void} [\phago_{n}]_{\zero})(id)
(\mset{P}_{\; \cell{\zero}{}}) = 0$, but again it is easy because the result
depends only on $\omega^{\sys}$, hence $\mu_{1} = \mu_{2}$. 

Now, we introduce the stochastic bisimulation for the Brane Calculus
as the stochastic bisimulation on the Markov kernel $(\P, \Pi,
\theta_{\sys})$.  We show that systems which are associated with the
same function by our SOS are bisimilar, and that the bisimulation
extends the structural equivalence.

Lemma~\ref{lm:uniqueness} shows that the operational semantics
induces a function $\vartheta \colon \P \to \Delta(\P, \Pi)^{\A^{+}_{\sys}}$
defined by
\begin{equation*}
  \vartheta(P) = \mu
  \qquad
  \text{ iff }
  \qquad
  P \to \mu
\end{equation*}

The next lemma shows that there is a relation between $\vartheta$
and the function $\theta_{\sys}$ that organizes $\P$ as a Markov kernel.
Indeed, it reflects the similarity between Definition~\ref{def:thetasystems}
and Definition~\ref{def:measop}.
\begin{lemma}
If $(\P, \Pi, \theta_{\sys})$ is the Markov kernel of system terms and
$\vartheta \colon \P \to \Delta(\P, \Pi)^{\A^{+}_{\sys}}$ is the function
induced by the SOS, then for any $P \in \P$, $\alpha \in \A^{+}_{\sys}$,
and $\mset{P} \in \Pi$,
\begin{equation*}
  \theta_{\sys}(\alpha)(P)(\mset{P}) = \vartheta(P)(\alpha)(\mset{P}).
\end{equation*}
\end{lemma}

A direct consequence of the previous lemma is that if our SOS
assigns to different systems the same function, then they are 
stochastic bisimilar with respect to the bisimulation on Markov 
processes.

\begin{corollary}
For arbitrary $P, Q \in \P$, if $P \to \mu$ and $Q \to \mu$, then
$P \sim_{(\P, \Pi, \theta_{\sys})} Q$.
\end{corollary}
This guarantees that we can safely define the stochastic bisimulation
for the Brane Calculus as the stochastic bisimulation on 
$(\P, \Pi, \theta_{\sys})$, as we do in the next definition.

\begin{definition}[Stochastic bisimulation on systems]
A \emph{rate-bisimulation relation} on systems in an equivalence
relation $\mathcal{R} \subseteq \P \times \P$ such that for arbitrary
$P, Q \in \P$ with $P \to \mu$ and $Q \to \mu'$,
\begin{equation*}
  (P, Q) \in \mathcal{R}
  \qquad
  \text{iff}
  \qquad
  \mu(\alpha)(C) = \mu'(\alpha)(C)
  \quad 
  \text{for any $C \in \Pi(\mathcal{R})$ and any $\alpha \in \A^{+}_{\sys}$}
\end{equation*}
Two systems $P, Q \in \P$ are stochastic bisimilar, written 
$P \approx Q$, iff there exists a rate bisimulation relation $\mathcal{R}$
such that $(P, Q) \in \mathcal{R}$.
\end{definition}

The next theorem provides a characterization of stochastic bisimulation
stating that $\approx$ is the smallest rate-bisimulation relation on $\P$.
\begin{theorem} \label{th:smallestbisim}
The stochastic bisimulation relation $\approx$ is the smallest equivalence
relation on $\P$ such that for arbitrary $P, Q \in \P$ with $P \to \mu$ 
and $Q \to \mu'$,
\begin{equation*}
  P \approx Q
  \qquad
  \text{iff}
  \qquad
  \mu(\alpha)(C) = \mu'(\alpha)(C)
  \quad 
  \text{for any $C \in \Pi(\approx)$ and any $\alpha \in \A^{+}_{\sys}$}.
\end{equation*}
\end{theorem}

\begin{example} \label{ex:stochbisims}
For arbitrary $\sigma \in \M$
\begin{enumerate}
  \item $\cell{\zero}{ \cell{\sigma}{ \void } } \approx \void$
  \item 
  $\cell{\pino_{n}(\sigma) | \pino_{n}(\sigma)}{ \void } \not\approx 
     \cell{\pino_{n}(\sigma).\pino_{n}(\sigma)}{ \void }$
  \item 
  $\cell{\cophago_{n}(\zero) | \coexo_{m}}{ \void } \approx
     \cell{\cophago_{n}(\zero).\coexo_{m}}{ \void } \approx
     \cell{\cophago_{n}(\zero)}{ \void }$
\end{enumerate} 
\end{example}
Examples \ref{ex:stochbisims}(1) and (3) are peculiar to the Brane
Calculus; (2) is interesting because in the semantics without
stochastic features the two processes are bisimilar:
$\cell{\pino_{n}(\sigma) | \pino_{n}(\sigma)}{ \void } \sim
\cell{\pino_{n}(\sigma).\pino_{n}(\sigma)}{ \void }$.

The next theorem shows that our stochastic bisimulation behaves
correctly with respect to structural equivalence;
it is a direct consequence of Lemma~\ref{lm:stransitionisuptoequiv} 
and Theorem~\ref{th:smallestbisim}.
\begin{theorem} \label{th:sbisimuptoequiv} 
For arbitrary $P, Q \in \P$, if $P \equiv Q$, then $P \approx Q$.
\end{theorem}

In addition to to the result of the previous theorem, notice that
$\approx$ is strictly larger then $\equiv$, indeed from 
Example~\ref{ex:stochbisims}(1) we have that, for any $\sigma \in \M$,
$\cell{\zero}{\cell{\sigma}{}} \approx \void$, however $\cell{\zero}{\cell{\sigma}{}} \not\equiv \void$.

\section{Conclusions}\label{sec:concl}
In this paper we have presented a stochastic extension of the Brane
Calculus; brane systems are interpreted as Markov processes over the
measurable space generated by terms up-to syntactic congruence, and
where the measures are indexed by the actions of Brane Calculus.  For
finding the correct actions, we have introduced a labelled transition
system for Brane Calculus.  Finally, we have provided a SOS
presentation of this stochastic semantics, which is compositional and
syntax-driven.

Stochastic semantics for calculi of biological compartments (but not
Brane Calculus) have been given in literature;
see~\cite{brodoDP:pact07,bv:QAPL09} for stochastic versions of
BioAmbients and \cite{versari-busi:cmsb07} for a stochastic
$\pi$-calculus with polyadic synchronisation.  However, these
semantics are ``pointwise'' and not structural, tailored for
stochastic simulations using Gillepie algorithm.  As shown in
Section~\ref{sec:bisim}, a ``pointwise'' semantics can be readily
obtained from the SOS given in this paper.  An interesting future work
is to investigate how these simulation algorithms and techniques can
be adapted to our setting.

There are several directions for further work.  First, we would like
to prove that the stochastic bisimilarity (which corresponds to Markov
bisimilarity) is a congruence.  Then, we want to extend the theory to
cover a notion of ``approximate behaviour'', in order to measure how
much two systems are bisimilar; this is important in biological
contexts, where usually we can compare only with approximate data
(e.g. coming from experiments).

We can consider also to add further constructs to the Brane Calculus,
like ``bind\&release'' and replication.  For the latter, we should add
rules like $\frac{P\circ !P \trs{\alpha} P'}{!P \trs{\alpha} P'}$ to
the LTS of Table~\ref{tbl:lts}; on the stochastic side, these rules
would lead to a new case in Definition~\ref{def:meta-theta}, which we
expect to be a fixed point equation to be solved in a suitable domain
(e.g., complete metric spaces).

Finally, we would like to apply the present approach to other
measurable aspects; in particular, geometric (e.g. volumes), physic
(e.g. pressure, temperature) and chemical aspects are of great interest
in the biological domain.

\paragraph{Acknoledgements} 
We would like to thank Radu Mardare for providing us an updated 
preliminary copy of \cite{cm:quest10} and for the valuable discussions
about the definition of Markov Processes.

\appendix
\section{Some measure theory}
\label{app:measuretheory}

Given a set $M$, a family $\Sigma$ of subsets of $M$ is called
a \emph{$\sigma$-algebra} if it contains $M$ and is closed under 
the formation of complements and (infinite) countable unions:
\begin{enumerate}
  \item $M \in \Sigma$;
  \item $A \in \Sigma$ implies $A^{c} \in \Sigma$, 
    where $A^{c} = M \setminus A$;
  \item $\set{A_i}_{i \in \N} \subset \Sigma$ implies 
    $\bigcup_{i \in \N} A_i \in \Sigma$.
\end{enumerate}
Since $M \in \Sigma$ and $M^{c} = \emptyset$, $\emptyset \in \Sigma$,
hence $\Sigma$ is nonempty by definition.
A $\sigma$-algebra is closed under countable set-theoretic operations:
is closed under finite unions ($A, B \in \Sigma$ implies $A \cup B = A \cup B \cup \emptyset \cup \emptyset \cup \dots \in \Sigma$), countable intersections (by DeMorgan's law $A \cap B = (A^{c} \cup B^{c})^{c}$ 
in its finite and inifite version), and countable subtractions 
($A, B \in \Sigma$ implies $A \setminus B = A \cap B^{c} \in \Sigma$).
\begin{definition}[Measurable Space]
Given a set $M$, and a $\sigma$-algebra on $M$, the tuple 
$(M, \Sigma)$ is called a \emph{measurable space}, the 
elements of $\Sigma$ \emph{measurable sets}, and $M$ the
\emph{support-set}.
\end{definition}

A set $\Omega \subseteq 2^{M}$ is a \emph{generator for the 
$\sigma$-algebra $\Sigma$} on $M$ if $\Sigma$ is the closure of 
$\Omega$ under complement and countable union; we write 
$\sigma(\Omega) = \Sigma$ and say that $\Sigma$ is generated by
$\Omega$. A generator $\Omega$ for $\Sigma$ is a \emph{base of 
$\Sigma$} if it has disjoin elements.
Note that the $\sigma$-algebra generated by a $\Omega$ is
also the smallest $\sigma$-algebra containing $\Omega$, that is,
the intersection of all $\sigma$-algebras that contain $\Omega$.
In particular it holds that a completely arbitrary intersection of 
$\sigma$-algebras is it self a $\sigma$-algebra.
A $\sigma$-algebra generated by $\Omega$, denoted by 
$\sigma(\Omega)$, is minimal in the sense that if $\Omega \subset \Sigma$
and $\Sigma$ is a $\sigma$-algebra, then 
$\sigma(\Omega) \subset \Sigma$.
Some facts about a generators are that, if $\Omega$ is a $\sigma$-algebra
then obviously $\sigma(\Omega) = \Omega$; if $\Omega$ is empty or 
$\Omega = \set{\emptyset}$, or $\Omega = \set{M}$, then 
$\sigma(\Omega) = \set{\emptyset, M}$; if $\Omega \subset \Sigma$ and 
$\Sigma$ is a $\sigma$-algebra, then $\sigma(\Omega) \subset \Sigma$.

A \emph{measure} on a measurable space $(M, \Sigma)$ is a function
$\mu \colon \Sigma \to \R^{+}$ such that
\begin{enumerate}
  \item $\mu(\emptyset) = 0$;
  \item for any disjoint sequence $\set{N_i}_{i \in I} \subseteq \Sigma$ with
  $I \subseteq \N$, it holds 
  $\mu(\bigcup_{i \in I} N_i) = \sum_{i \in I} \mu(N_i)$.
\end{enumerate}
The triple $(M, \Sigma, \mu)$ is called a \emph{measure space}.
A measure space $(M, \Sigma, \mu)$ is called \emph{finite} if 
$\mu(M)$ is a finite real number; it is called $\sigma$-finite if $M$ can 
be decomposed into a countable union of measurable sets of finite measure. A set in a measure space has \emph{$\sigma$-finite measure} 
if it is a countable union of sets with finite measure.
Specifying a measure includes specifying its domain. If $\mu$ is a 
measure on a measurable space $(M, \Sigma)$ and $\Sigma'$ is
a $\sigma$-field contained in $\Sigma$, then the restriction $\mu'$ of 
$\mu$ to $\Sigma'$ is also a measure, and in particular a measure
on $(M', \Sigma')$, for some $M' \subseteq M$ for which $\Sigma'$ is
a $\sigma$-algebra on $M'$.
\\
A notable measure is the \emph{Dirac measure}. 
If $\Omega$ is a base for $(M, \Sigma)$, $N \in \Omega$ and $r \in \R^{+}$,
then the function $f \colon \Omega \to \R^{+}$
\begin{equation*}
f(N') = 
\begin{cases}
  r & \text{if $N' = N$} \\
  0 & \text{if $N' \neq N$}
\end{cases}
\end{equation*}
can be extended, by $f(\bigcup_{i \in I} N_i) = \sum_{i \in I} f(N_i)$, to a
measure on $(M, \Sigma)$ denoted by $D(r, N)$ and called the 
\emph{$r$-Dirac measure on $N$}.

Let $\Delta(M,\Sigma)$ be the class of measures on $(M, \Sigma)$.
It can be organized as a measurable space by considering the
$\sigma$-algebra generated by the sets 
$N_{S, r} = \set{\mu \in \Delta(M, \Sigma) : \mu(S) \geq r}$, for arbitrary
$S \in \Sigma$ and $r > 0$ (the support-set is $\Delta(M,\Sigma)$ and
the $\sigma$-algebra is $\sigma(\bigcup_{S \in \Sigma, r \in \R^{+}} N_{S,r})$).

Given two measurable spaces $(M, \Sigma)$ and $(N, \Theta)$ a mapping
$f \colon M \to N$ is \emph{measurable} if for any $T \in \Theta$, 
$f^{-1}(T) \in \Sigma$. Measurable functions are closed under composition:
given $f\colon M \to N$ and $g\colon N \to O$ measurable functions then
$g \circ f \colon M \to O$ is also measurable.

\end{document}